%% file: EVN_gk.tex
\documentclass[cits]{PoS}

\input{journals.tex}

\usepackage{amsmath}
\usepackage{dcolumn}

\makeatletter
\newcommand\tabcaption{\def\@captype{table}\caption}
\makeatother

\title{Effects of the turbulent ISM on radio observations of quasars}

\ShortTitle{Effects of the turbulent ISM on radio observations of quasars}

\author{\speaker{Krisztina \'Eva Gab\'anyi} \\%
        Max-Planck-Institut f\"ur Radioastronomie, Bonn, Germany\\
        E-mail: \email{gabanyik@mpifr-bonn.mpg.de}}

\author{S. Britzen, T.~P. Krichbaum, U. Bach, L. Fuhrmann, A. Kraus, A. Witzel and J.~A. Zensus\\
 Max-Planck-Institut f\"ur Radioastronomie, Bonn, Germany\\
E-mail: \email{sbritzen@mpifr-bonn.mpg.de}, \email{tkrichbaum@mpifr-bonn.mpg.de}, \email{ubach@mpifr-bonn.mpg.de}, \email{fuhrmann@mpifr-bonn.mpg.de}, \email{akraus@mpifr-bonn.mpg.de}, \email{awitzel@mpifr-bonn.mpg.de}, \email{azensus@mpifr-bonn.mpg.de}}

\abstract{In radio bands, the study of compact radio sources can be affected by propagation effects
through the Interstellar Medium. These are usually attributed to the presence of 
turbulent intervening plasma along the line of sight. In this talk, 
two of such effects are presented. The line of sight of B\,2005+403 
passes through the heavily scattered region of Cygnus, which causes 
substantial angular broadening of the source images obtained at frequencies between 0.6\,GHz and 
8\,GHz. At higher frequencies, however the intrinsic source structure shines 
through. Therefore, multi-frequency VLBI observations allow to study the 
characteristics of the intervening material, the source morphology and the 
"coupling" of them in forming the observed image.

This article is based upon the published paper of Gab\'anyi et al. For more details see \cite{cikk}.}

\FullConference{8th European VLBI Network Symposium\\
                 September 26-29, 2006\\
                 Toru\'n, Poland}

\begin{document}

\section{Introduction}

The propagation of radio waves through the ionized interstellar medium causes
several effects, such as Faraday rotation and depolarization of polarized emission, dispersion of pulsar signals, scatter broadening of compact radio sources and intensity fluctuations caused by diffractive (DISS) and refractive (RISS) interstellar scintillation (ISS).
The detailed study of scatter broadening in compact radio sources can lead
to a better understanding of the interstellar medium.

The other prominent propagation effect addressed here is the so-called
Intraday Variability (IDV, \cite{idvh}), and the question of how it 
is related to the interstellar scintillation. IDV surveys show that a large fraction (up to 30\,\%) of all compact
flat spectrum radio sources show this effect, which is characterized by
variability amplitudes of up to $20-30$\,\% and variability timescales
ranging from less than one hour to several days (e.g. \cite{reduc_idv}, \cite{lovell} and references therein). 
If interpreted via source intrinsic incoherent emission processes, such short variability timescales imply -- via the light travel time argument --  apparent brightness temperatures of $T_\text{B} = 10^{16}\text{\,K} - 10^{19}$\,K, far in excess of the inverse-Compton limit (\cite{compton}).
With the assumption of relativistic Doppler-boosting the brightness temperatures can be reduced, which however would imply uncomfortably large Doppler boosting factors of $\delta \simeq 20 - 200$.

Alternatively, IDV in the radio-bands is also explained
extrinsically via scintillation of radio waves in the turbulent interstellar medium (ISM) of our Galaxy (e.g. \cite{gbi2}). 
The main problem in this interpretation is that it cannot explain the observed radio-optical broad-band correlations seen in at least some sources
(\cite{corr_var}). The recent detection of diffractive ISS in the source
J\,1819+3845 leads to micro-arcsecond source sizes and brightness temperatures of $\sim 10^{14}$\,K, which again 
require Doppler boosting factors of $\delta \simeq 100$ (\cite{diff_1819}).
It is therefore likely that the IDV phenomenon involves both, a combination of
source intrinsic {\it and} extrinsic effects (e.g. \cite{krich}).

We present measurements for the scatter broadened quasar B\,2005+403, combining Very Long Baseline Interferometry (VLBI) and flux density variability measurements obtained at various frequencies. The scatter broadening observed in the VLBI images at the lower frequencies is measured and yields further constraints to the properties of the intervening ISM. Additional parameters for the ISM are obtained from the IDV monitoring of the source, performed with the Effelsberg 100\,m telescope.

B\,2005+403 is a flat spectrum quasar  
($\alpha_{\rm 0.3/5\,GHz} = 0.3$, with $ S \sim \nu^{\alpha}$; \cite{spekt})
at a redshift of $z=1.736$ (\cite{redshift}). It is located close to the Galactic plane at 
$l$\,=\,$76.82^\circ$, $b$\,=\,$4.29^\circ$ behind the Cygnus super-bubble region. Earlier studies 
showed that interstellar scattering affects the VLBI image of the source, causing 
angular broadening at frequencies below 5\,GHz (see \cite{desai} and references therein). 
The high value of the scattering measure ($ \text{SM}$\,=\,$\int_0^L C^{2}_{N}\left( s \right) ds $ \footnote{The Scattering 
Measure is the path integral over the coefficient $C^{2}_{N}$ of the electron density fluctuation wavenumber spectrum.}), 
derived for the line of sight of B\,2005+403 by \cite{fey}, reflects the strong influence of the interstellar medium.

\section{VLBI data and data reduction}

The VLBI observations of B\,2005+403 were performed at frequencies between
 $1.6$\,GHz and $43$\,GHz. 

After correlation at the VLBI correlators in Socorro (NRAO) or Bonn (MPIfR),
a priori amplitude calibration using system temperature measurement and
fringe-fitting were performed using the standard AIPS
analysis tasks.
Editing, phase and amplitude self-calibration and imaging were performed using AIPS and the Caltech DIFMAP packages. 

To study the scattering effects following \cite{fey}, we measured the total angular 
source size as a function of observing frequency by fitting the visibility data of 
B\,2005+403 at all frequencies with \emph{one} Gaussian component within DIFMAP. 

Additional size measurements of B\,2005+403 were used from \cite{fey} and \cite{desai}.

\section{Propagation effect I: Scatter broadening of B\,2005+403}

\begin{figure}
\begin{minipage}[b]{0.45\textwidth}
\centering
\includegraphics[width=7cm]{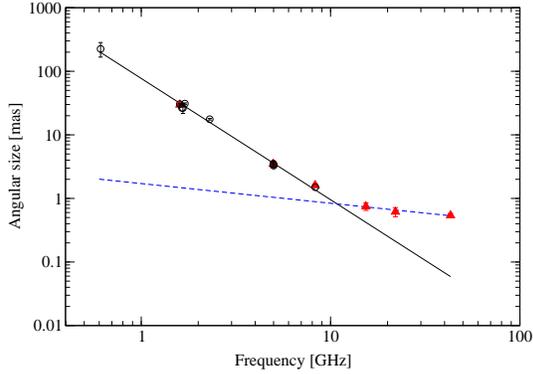}
 \caption{The measured angular size plotted versus observing frequency. The solid black line represents a power law fit to the data in the range from 0.6\,GHz to 8\,GHz. The dashed blue line represents a power law fitted to the data in the range from 15\,GHz to 43\,GHz. Black circles denote data from the literature (\cite{fey}, \cite{desai}), red triangles denote our data.}
 \label{fig:logplot}
\end{minipage}
\hfill
\begin{minipage}[b]{0.45\textwidth}
\begin{tabular}{*1{>{$}c<{$}}c D{.}{.}{1} @{ $\pm$ } D{.}{.}{1} c}
\hline
\hline
\nu & Epoch & \multicolumn{2}{c}{$\theta$} & Ref.\\ 
\text{[GHz]} & & \multicolumn{2}{c}{[$\text{mas}$]} & \\
\hline 
0.6 & Oct 1986 & 225.0&58.0 & \cite{fey} \\
1.6 & 1998.14 & 29.8&0.5 & \cite{cikk} \\
1.7 & Mar 1986 & 26.4&4.7 & \cite{fey} \\
1.7 & Jan 1997 & 31.0&0.8 & \cite{desai} \\
2.3 & Jan 1997 & 17.6&0.5 & \cite{desai} \\
5 & Oct 1985 & 3.5&0.4 & \cite{fey} \\
5 & 1996.82 & 3.5&0.1 & \cite{cikk} \\
5 & Jan 1997 & 3.4&0.1 & \cite{desai} \\
8 & 1996.83 & 1.6&0.1 & \cite{cikk} \\
15 & 1996.73 & 0.8&0.1 & \cite{cikk} \\
22 & 1996.73 & 0.6&0.1 & \cite{cikk} \\
43 & 1996.73 & 0.5&0.1 & \cite{cikk} \\
\hline
\end{tabular}
\tabcaption{\label{tab:scat}The measured angular sizes of B\,2005+403 at different frequencies.}
\end{minipage}
\end{figure}

The angular source size of B\,2005+403 versus frequency is plotted in Fig. \ref{fig:logplot}, using our data 
and including the previous measurements from the literature. The corresponding values are given
in Table \ref{tab:scat}. To quantify the scatter broadening, which dominates at lower frequencies, 
a power law was fitted to the size-frequency relation in the frequency range from $0.67$\,GHz to $8$\,GHz, yielding
$\theta=(77.1 \pm 4.0)\cdot (\nu/1\text{\,GHz})^{-(1.91 \pm 0.05)} \text{\,mas}$.
This fit is shown as the black line in Figure \ref{fig:logplot}. (Excluding the 
data point at 8\,GHz does not change the slope significantly,
yielding: $\theta =(77.7 \pm 4.0) \cdot (\nu/1\text{\,GHz})^{-1.92 \pm 0.06}$.) 
Above 8\,GHz, the extrapolated scattering size becomes smaller than the measured source size. 
This indicates that towards higher frequencies scattering effects are less dominant and that
the intrinsic structure of the sources shines through. The differences between the 
extrapolated scattering size and the measured source size at all three frequencies (15\,GHz, 22\,GHz, and 43\,GHz) are
of the order of 0.4\,mas. 

To characterize the frequency dependence of the intrinsic source size, 
a power law was fitted to the size-frequency relation above 8\,GHz, yielding:
$\theta_\text{int}=(1.7\pm 0.4)\cdot (\nu/1\text{\,GHz})^{-(0.31 \pm 0.07)} \text{\,mas}$.
The dashed blue line in Fig. \ref{fig:logplot} shows this fit. With this line, it was possible to
obtain an upper limit on the intrinsic source size at lower frequencies where direct size
measurements with VLBI are not possible. Hence a lower limit on the brightness temperature could also be calculated.
Below 8\,GHz, the intrinsic source sizes are in the range of $1 - 1.5$\,mas. 
The corresponding lower limits to the brightness temperatures at 1.6\,GHz, 5\,GHz and 8\,GHz 
are $T_\text{B}\ge(0.6\pm0.5)\cdot10^{12}$\,K, 
$T_\text{B}\ge(0.14\pm0.09)\cdot10^{12}$\,K, and $T_\text{B}\ge(0.08\pm0.04)\cdot10^{12}$\,K, respectively. These numbers
are in accordance with typical brightness temperatures measured with VLBI 
and neither strongly violate the inverse-Compton limit nor do they indicate excessive Doppler-boosting.

The size of the scattering disk at a given frequency can be estimated via the deconvolution formula:
$\theta_{\rm scat} = \sqrt{\theta_{\rm obs}^2 - \theta_{\rm int}^2}$. From the power law 
fit, the scattering size at 1\,GHz
$\theta_{\text{1\,GHz}}$ is $(77.1 \pm 4.0)$\,mas. Following \cite{tc93} and assuming a Kolmogorov turbulence,
the scattering measure can be calculated.
The derived SM of $(0.43 \pm 0.04) \text{m}^{-20/3} \text{\,kpc}$ is in good agreement and consistent with the
previous measurement of \cite{fey}, though with much improved accuracy. 

\subsection{Propagation effect II: IDV behaviour of B\,2005+403} \label{short}

\begin{table}
\centering
\begin{tabular}{*6{>{$}c<{$}}}
\hline
\hline
\text{Source name} & \langle S\rangle \text{\,[Jy]} & \sigma \text{\,[Jy]} & m \text{\,[\%]} & Y \text{\,[\%]} & \chi^2_\text{r} \\ 
\hline
\multicolumn{6}{l}{\(\nu=1.6\)\,GHz, \(m_0=0.25\)\,\%} \\
\hline
\text{B\,2005+403} & 2.430 & 0.024 & 1.01 & 2.93 & 5.593 \\
\text{NGC\,7027} & 1.906 & 0.004 & 0.19 & 0 & 0.197 \\
\text{B\,2021+614} & 2.193 & 0.006 & 0.25 & 0 & 0.365 \\
\hline
\hline
\multicolumn{6}{l}{\(\nu=5\)\,GHz, \(m_0=0.20\)\,\%}\\
\hline 
\text{B\,2005+403} & 2.905 & 0.013 & 0.45 & 1.18 & 3.107 \\
\text{NGC\,7027} & 5.489 & 0.011 & 0.20 & 0 & 0.621 \\
\hline
\end{tabular}
\caption{\label{tab:m_y}The variability parameters of B\,2005+403 and the secondary calibrators
at 1.6\,GHz (top) and 5\,GHz (bottom). Col. 2 shows the mean flux density in Jy, col. 3 its standard deviation, col. 4 the 
modulation index, col. 5 the noise-bias corrected variability amplitude and col. 6 the reduced $\chi^2_\text{r}$. For a detailed definition of these values see \cite{reduc_idv}.}
\end{table}

In November and December 2003 the flux-density variability of B\,2005+403 was monitored 
with the Effelsberg 100 meter radio telescope with high time resolution. 
In Table \ref{tab:m_y}, the results from these measurements are summarized:
To characterize the variability amplitudes and their significance, we follow the methods described in \cite{reduc_idv}.
Column 2 of the table gives the average flux density, col. 3 the rms standard deviation,  col. 4 the modulation
index ($m=100\cdot\sigma/\langle S \rangle$), col. 5 the variability 
amplitude (defined as  $Y = 3 \sqrt{m^2-m_{0}^2}$, where $m_0$ 
is the modulation index of a source regarded to be stationary during the observation)
and col. 6 the reduced $\chi^2$. A value of $Y$ is only given, if the $\chi^2-$test gives a higher than 99.9\,%
probability for significant variations. 

The variability amplitude of B\,2005+403 decreases with frequency. Formally we measure a modulation index of $m=1.01$\,\%
at 1.67\,GHz and of $m=0.45$\,\% at 5\,GHz. At this frequency the detection of IDV is marginal. We note that
the model of refractive interstellar scintillation in the weak regime predicts a decrease of the modulation index
with increasing frequency (\cite{gbi2}). This is consistent with our observations.

The source size and the apparent brightness temperature can be determined 
from the interstellar scintillation model, assuming that the main reason for the observed
variability is the motion of the Earth through the scintillation pattern.
In this scenario the variability time-scale in days is (\cite{gbi2}): 
$t=18\cdot\theta_\text{scat}\cdot L_\text{kpc}\cdot\left(\frac{v}{50 \text{\,km s}^{-1}}\right)^{-1} \label{scint_time}$.

Cordes and Lazio modeled the electron density turbulences of the Milky Way (NE2001, \cite{gal_model}). They
assumed a distance of 2.35\,kpc to the scattering ``clump'' responsible for the angular broadening of
B\,2005+403. The characteristic variability time-scales derived from the light curves is less than a day in B\,2005+403.
A screen located at kpc distance cannot explain the observed
short variability time-scale. With a scattering size of $ \theta_\text{scat} = 30$\,mas at 1.6\,GHz, a screen distance
of 2.35\,kpc and typical Galactic velocities of $\leq 220$\,km/s, one would expect to see variations
on time-scales $\geq 288$ days. 

To reproduce a variability time-scale of $\leq 1$\,day at 1.6\,GHz within the aforementioned model,
the following constraints for the scattering size and the screen distance are obtained: $\theta_\text{eff} L \leq 2.2 \cdot 10^{-2}$\,mas kpc.
For the measured scattering size of $\sim 30$\,mas, this leads to an unreasonably nearby screen of $\leq 1$\,pc.
The only way out of this dilemma is a smaller scattering size. Adopting a minimum screen distance
of at least 10\,pc as required for the interpretation of the ultra fast scintillators (\cite{diff_1819} and references therein), 
one obtains $\theta_\text{scat} \leq 2.2$\,mas. 

An upper limit on the distance of the screen is obtained from the restriction of the source size
via the inverse-Compton limit of $T_\text{B}^\text{IC}=10^{12}$\,K. The requirement that the brightness temperature be lower than 
this leads to a source size of 
$\theta_\text{int} = \sqrt{1.77 \cdot 10^{12} S_{\nu} \nu^{-2} T_\text{B}^{-1} \delta^{-1} (1+z)} \geq 0.6\text{\,mas}$, 
where $S_{\rm 1.6\,GHz} = 2.4$\,Jy is adopted from Table \ref{tab:m_y} and $\delta \simeq 10$ for the Doppler factor and
$\nu$ is in the units of GHz.
Approximating $\theta_\text{scat}$ with this size and a relative screen velocity in the range 
$(50 \text{ to }220)$\,km/s, leads to an upper limit of the screen distance in the range $(9 \text{ to } 41)$\,pc.

\section{Conclusion}

The observed IDV in B\,2005+403 can be explained with a screen located at a distance less than or equal to 
$41$\,pc that can be characterized by a scattering size of $0.6\text{\,mas} \leq \theta_\text{scat} \leq 2.2$\,mas. 
The corresponding scattering measure at 1.6\,GHz then is in the order of 
$10^{-3}\text{\,m}^{-20/3}\text{\,kpc}$.
These values are considerably lower than the scattering measure of the more distant screen, which is thought to be responsible for
the scatter broadening in B\,2005+403. Therefore the observed IDV is not caused by this distant screen.
Most likely, one has to assume multiple scattering by at least two spatially and 
physically very different plasma screens. In this scenario,
the first screen leads to a significant scatter broadening 
of the source image, which then by the second screen leads to only very weak scattering due to large quenching effects ($\theta_\text{source}
>> \theta_\text{scat}$). This quenched scattering (\cite{gbi2}) by the second screen can explain the
relatively low variability amplitudes of $\leq 1$\,\%, observed in the IDV experiments at 5\,GHz and 1.6\,GHz. 
Quantitatively, this can be verified using equation (20) of \cite{goodman}, which relates the variability index, the
scattering measure, the effective source size and the screen distance. With the parameters from above,
a modulation index of $m \leq 1$\,\% is obtained, in good agreement with the observations.

\end{document}

%% file: journals.tex
\def\aj{AJ}%
\def\apj{ApJ}%
\def\apjl{ApJ}%
\def\apjs{ApJS}%
\def\aap{A\&A}%
\def\mnras{MNRAS}%